\documentclass[aps,prl,twocolumn]{revtex4}
\usepackage[dvips]{graphicx}
\def\gsim{\buildrel {\textstyle >}\over {_\sim}}
\def\lsim{\buildrel {\textstyle <}\over {_\sim}}
\begin{document}
\title{
Vortex-induced topological transition of the bilinear-biquadratic Heisenberg antiferromagnet on the triangular lattice
}

\author{
Hikaru Kawamura and Atsushi Yamamoto
}

\affiliation{Department of Earth and Space Science, Faculty of Science,
Osaka University, Toyonaka 560-0043,
Japan}
\date{\today}
\begin{abstract}
The ordering of the classical Heisenberg antiferromagnet on the triangular lattice with the the bilinear-biquadratic interaction is studied by Monte Carlo simulations. It is shown that the model exhibits a topological phase transition at a finite-temperature driven by topologically stable vortices, while the spin correlation length remains finite even at and below the transition point. The relevant vortices could be of three different types, depending on the value of the biquadratic coupling. Implications to recent experiments on the triangular antiferromagnet NiGa$_2$S$_4$ is discussed.
\end{abstract}
\maketitle

Antiferromagnetic (AF) Heisenberg model on the two-dimensional (2D) triangular lattice has been studied extensively as a typical example of geometrically frustrated magnets. Inspired by recent experiments on a variety of triangular magnets, including 
% $^3$He film \cite{Hukuyama}, organic magnets \cite{Kanoda}, 
NiGa$_2$S$_4$\cite{Nakatsuji,Nambu} and NaCrO$_2$\cite{Olariu}, renewed interest has now arisen in this model. The triangular-lattice Heisenberg antiferromagnet (AF) with the nearest-neighbor bilinear exchange is known to exhibit a magnetic long-range order (LRO) at $T=0$, the so-called 120$^\circ$ structure, in either case of quantum $S=1/2$ or classical $S=\infty$ spin. Because of the two-dimensionality of the lattice, the AF LRO is established only at $T=0$, while the associated  spin correlation length diverges exponentially toward $T=0$. 

Some time ago,  it was demonstrated by Kawamura and Miyashita (KM) that the triangular Heisenberg AF bears a topologically stable point defect characterized by a two-valued topological quantum number, $Z_2$ vortex, in contrast to its unfrustrated counterpart \cite{KawaMiya}. Existence of such a vortex has become possible owing to the noncollinear nature of the spin order induced by frustration. KM suggested  that  the triangular Heisenberg AF might exhibit a genuine thermodynamic transition at a finite temperature associated with the condensation (binding-unbinding) of $Z_2$ vortices. This topological transition is of different character from the standard Kosterlitz-Thouless (KT) transition in that the spin correlation length does not diverge even at and below the transition point and the spin correlation in the low-temperature phase decays exponentially. The topological transition occurs between the two spin {\it paramagnetic\/} states. 

On experimental side, recent data by Nakatsuji {\it et al\/} on the S=1 triangular Heisenberg AF NiGa$_2$S$_4$ are of particular interest: While no magnetic LRO is observed to low temperature, the low-temperature specific heat exhibits a $T^2$ behavior, suggesting the existence of Goldstone modes associated with a broken continuous symmetry. Meanwhile, neutron scattering measurements suggested that the spin correlation length stayed short even at low temperature \cite{Nakatsuji}. To account for such peculiar experimental results, Tusnetsugu and Arikawa \cite{Tsunetsugu}, L\"auchli {\it et al\/} \cite{Lauchli} and Bhattacharjee {\it et al\/} \cite{Bhatta} proposed a scenario where the spin nematic order, either ferroquadratic (FQ) or antiferroquadratic (AFQ), play a dominant role. Their theoretical anlyses  were performed on the basis of the AF $S=1$ Heisenberg model with the blinear-biquadratic exchange. Experimentally, a weak but clear anomaly, possibly originated from some kind of phase transition, is observed in the susceptibility at $T\simeq 8.5$K \cite{Nakatsuji}.
%, although it is not entirely clear whether this anomaly is of either intrinsic or extrinsic origin. 
In the present letter, we address the issue of the nature of the experimentally observed  transition-like anomaly of NiGa$_2$S$_4$. 
%We propose that the observed ``transition'' might be originated from a vortex-induced topological transition inherent to the triangular Heisenberg AF.

The model considered is the $S=\infty$ version of the $S=1$ Hamiltonian used in Refs.\cite{Tsunetsugu,Lauchli,Bhatta}, {\it i.e.\/}, a classical Heisenberg AF on the 2D triangular lattice with the bilinear-biquadratic exchange described by
\begin{equation}
{\cal H} = -J \sum _{<ij>} \vec S_i\cdot \vec S_j - K \sum _{<ij>}(\vec S_i\cdot \vec S_j)^2, 
\end{equation}
where $J<0$ is the antiferromagnetic bilinear exchange, $K$ the biquadratic exchange which is  either FQ ($K>0$) or AFQ ($K<0$), and the sum is taken over all nearest-neighbor pairs on the lattice. While the biquadratic term is essential in stabilizing a hypothetical spin nematic order, its significance in real systems has not been established yet. The biquadratic term is usually small, while it has been argued that it could be large near the Mott transition or due to the effect of orbitals \cite{Lauchli}. In the present letter, following Refs.\cite{Tsunetsugu,Lauchli,Bhatta}, we assume (1), and investigate its finite-temperature ordering properties by means of Monte Carlo (MC) simulations.
%, paying particular attention to its topological defect structure. We deal with (A) the pure biinear case $K=0$, (B) the FQ case $K>0$, and (C) the AFQ case $K<0$, respectively. 

MC simulations are performed based on the standard heat-bath method. The system studied is of size $L\times L$,  $L$ being in the range from 48 to 192, with periodic boundary conditions. The system is gradually cooled from the high temperature, each run containing ($3\sim 6$)$\times $10$^5$ Monte Carlo steps per spin (MCS) at each temperature. Averages are then made over $5\sim 10$ independent runs.

\medskip\noindent
{\it Pure bilinear case $K=0$\/}

In the case of the bilinear interaction only ($K=0$), the ordering property of the model were studied extensively \cite{KawaMiya,Southern,Wintel}. 
%The ground state is the 120$^\circ$ spin structure with a three-sublattice periodicity. 
Numerical studies suggested that the model exhibited a $Z_2$ vortex-induced topological transition at $T=T_V\simeq 0.3$ (in units of $|J|$), at which the spin correlation length remains finite. The specific peak exhibits a rounded peak above $T_V$, while no appreciable anomaly is observed at $T_V$. The transition manifests itself as a dynamical anomaly \cite{KawaMiya,Ajiro}.

A convenient quantity characterizing such a vortex transition might be the {\it vorticity modulus\/}, which measures the stiffness of the system against spin deformation corresponding to vortex formation \cite{KawaKiku}. The vorticity modulus is defined by $v=\Delta F/\ln L$ where $\Delta F$ is the free-energy cost due to the introduction of an isolated vortex into the system. In MC simulations, $v$ can be calculated from appropriately defined fluctuations \cite{Southern}. In the vortex-unbounded phase, the system does not exhibit macroscopic stiffness against vortex formation with $v=0$, while, in the vortex-bounded phase, the system becomes stiff against vortex formation and $v>0$. 

%Note that in the triangular Heisenberg AF the vorticity modulus might behave differently from the helicity modulus: The latter should vanish even at $T<T_V$ in contrast to the vorticity modulus. This might be contrasted with  the familiar Kosterlitz-Thouless (KT) phase realized in the two-dimensional {\it XY} model, where both the helicity modulus and the vorticity modulus are nonzero. 

Our MC result of the vorticity modulus is shown in Fig.1a. The data indicate the occurrence of a vortex-induced topological transition at $T\simeq 0.28$, consistently with the previous results \cite{KawaMiya,Southern,Wintel}.

\medskip\noindent
{\it Ferroquadrapolar case $K>0$\/}

Next, we analyze the FQ case with $K>0$. The ground-state of three spins on a triangle is the 120$^\circ$ structure for $K<2/9$ (measured in units of $|J|$), whereas at $K=2/9$ it exhibits a discontinuous change into the collinear state with up-up-down (down-down-up) state as illustrated in Fig.2a, which remains to be the ground state up to $K=\infty$. Such a collinear ground state resembles the one of the triangular Ising AF, although in the present Heisenberg case the axis of spin collinearity can be arbitrary. In the collinear ground state, whether each spin points either up or down is not uniquely determined due to the frustration-induced local degeneracy: See Fig.2a.  Such a local degeneracy leads to a macroscopic degeneracy in an infinite triangular lattice. Indeed, one sees from exact information about the corresponding Ising model that the collinear ground state does not possess a true AF LRO, but only a quasi-LRO with power-law-decaying spin correlations \cite{Ising}. Meanwhile, since spins are aligned all parallel or antiparallel selecting a unique axis in spin space, the collinear ground state is characterized by the FQ LRO. The order parameter of the FQ state is a director, rather than the spin itself. In terms of a local quadrapole variable, $q_{i\mu\nu}=S_{i\mu}S_{i\nu}-(1/3)\delta_{\mu \nu}$, the FQ order parameter $Q_F$ might be defined by,
\begin{figure}[ht]
\begin{center}
\includegraphics[scale=0.5,angle=-90]{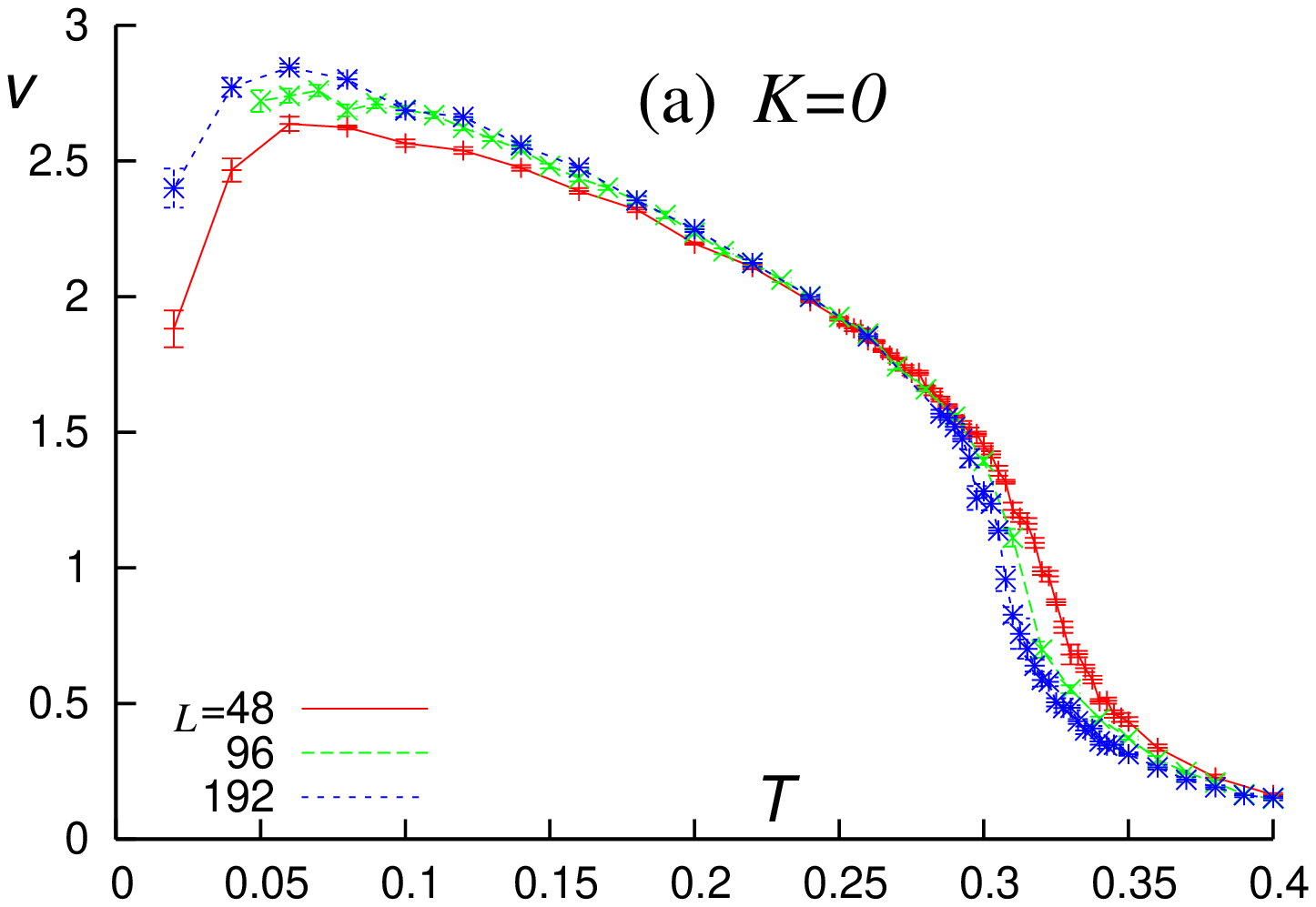}
\includegraphics[scale=0.5,angle=-90]{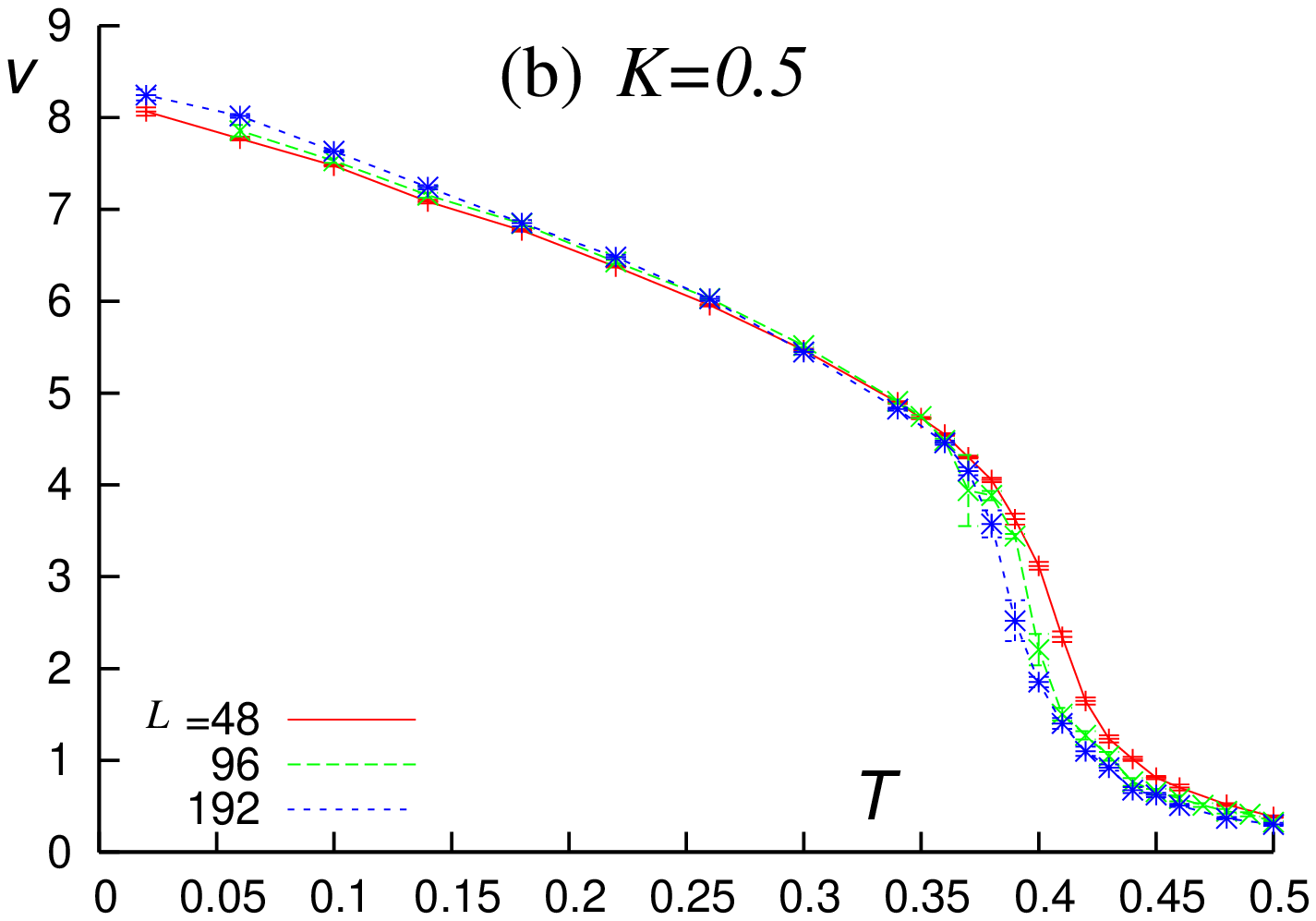}
\includegraphics[scale=0.5,angle=-90]{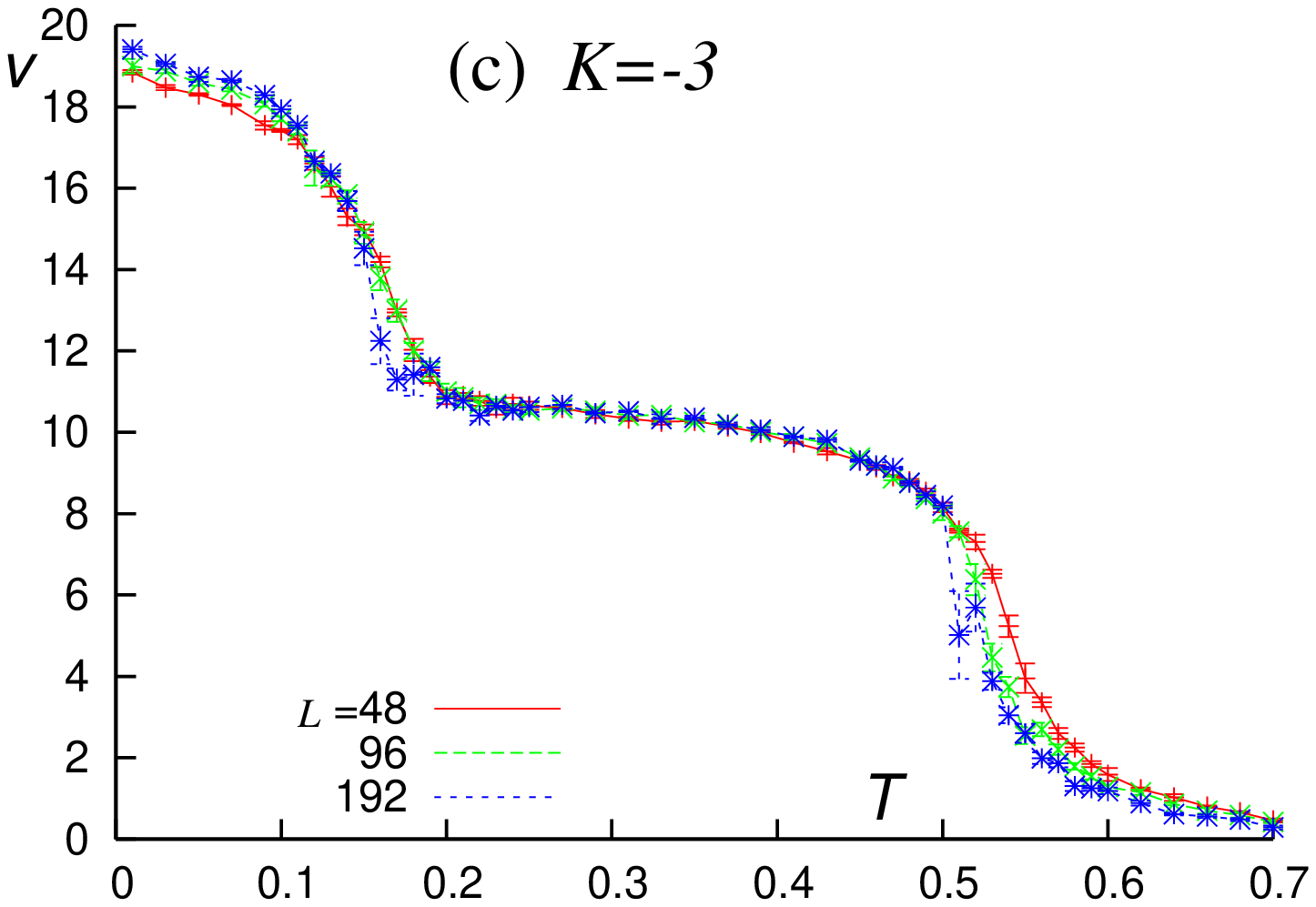}
\end{center}
\caption{
The temperature and size dependence of the vorticity modulus for (a) $K=0$, (b) $K=0.5$, and (c) $K=-3$.
}
\end{figure}
\begin{equation}
(Q_{F})^2=\frac{3}{2}\sum_{\mu,\nu =x,y,z}<(\frac{1}{N}\sum_iq_{i\mu\nu})^2>,
\end{equation}
where $<\cdots >$ represents a thermal average. 
%It is normalized to give unity for the collinear state.

In Fig.3a, we show for the case of $K=0.5$ the temperature dependence of $Q_F$ together with that of the Fourier magnetization $m_f$ defined by
\begin{equation}
(m_f)^2 = 2<|\vec m(\vec q)|^2>,\ \ \ \vec m(\vec q) = \frac{1}{N}\sum_i 
\vec S_ie^{i\vec q\cdot r_i},
\end{equation}
where $\vec q=(\frac{4\pi}{3},0)$. 
%It is normalized to give unity for the 120$^\circ$ structure. 
Although both  $Q_F$ and $m_f$ vanish in the thermodynamic limit at any $T>0$, one can still get useful information about the short-range order (SRO) from the corresponding finite-size quantities. As can be seen from Fig.3a, the FQ SRO develops rather sharply at $T\simeq 0.4$, while the standard AF SRO is kept smaller.
\begin{figure}[ht]
\begin{center}
\includegraphics[scale=0.55]{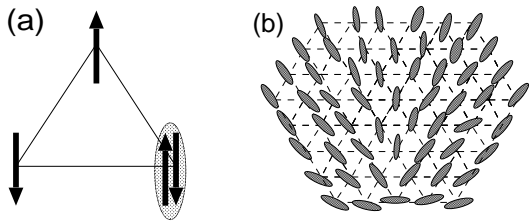}
\end{center}
\caption{
(a) Frustrated spins on a triangle in the FQ state. (b) $Z_2$ vortex formed by directors in the FQ state.
}
\end{figure}
\begin{figure}[ht]
\begin{center}
\includegraphics[scale=0.5,angle=-90]{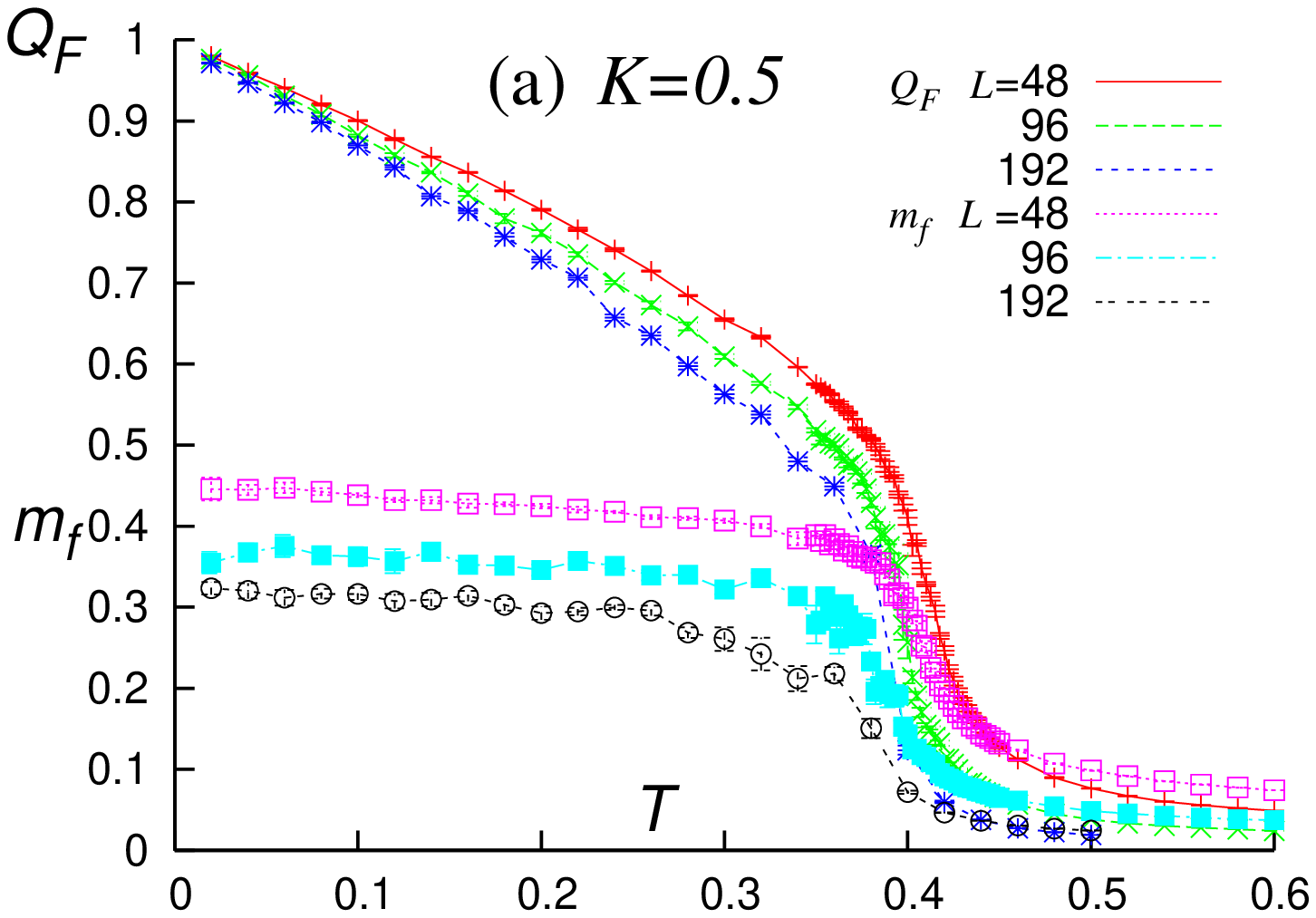}
\includegraphics[scale=0.5,angle=-90]{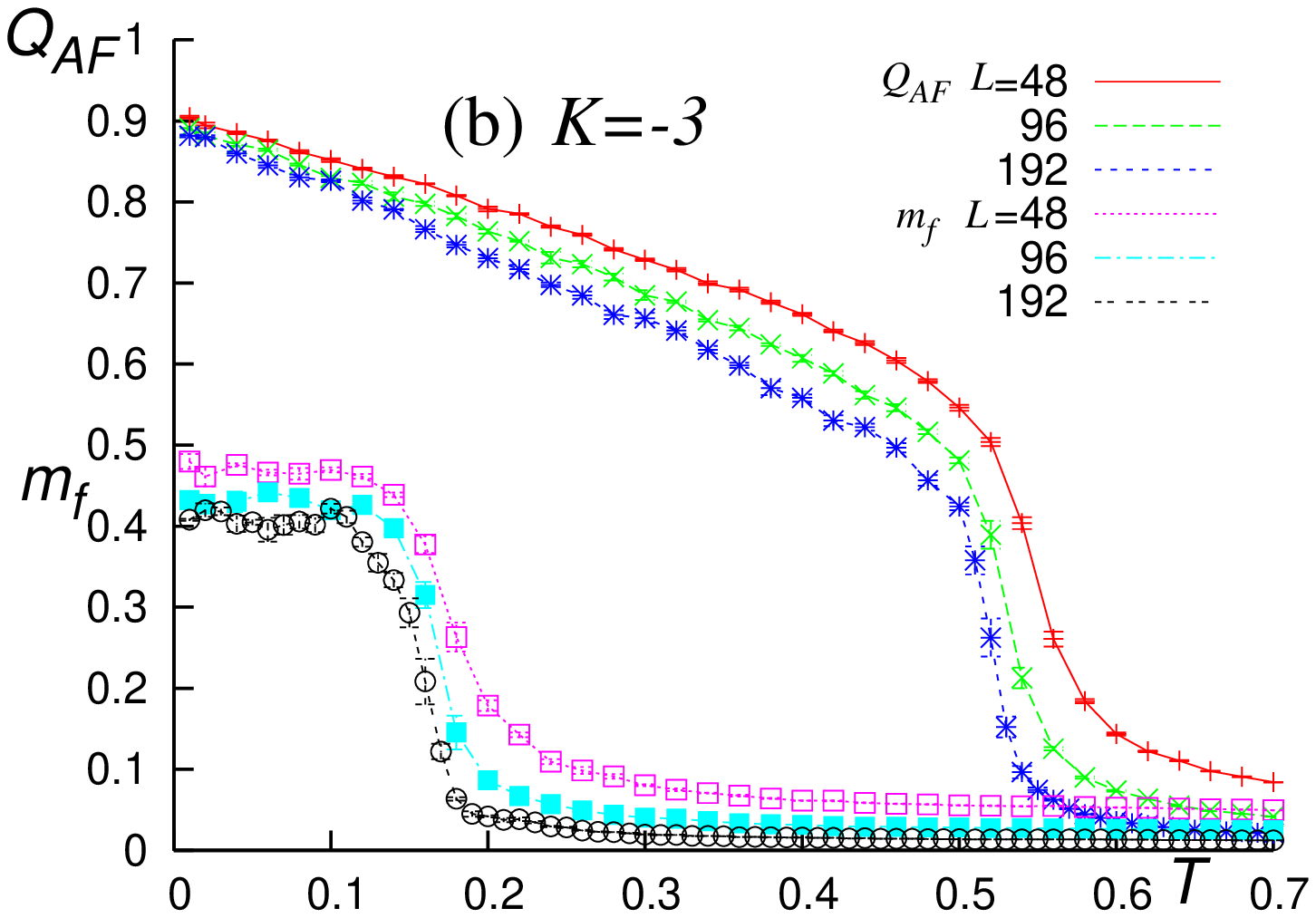}
\end{center}
\caption{
The temperature and size dependence of the FQ and AFQ order parameters, $Q_F$ and $Q_{AF}$, and the Fourier magnetization $m_f$ for the cases of (a) $K=0.5$ and (b) $K=-3$.
}
\end{figure}
The $Z_2$ vortex based on the noncollinear spin order is expected to survive at least up to $K=2/9$. Different situation, however, is expected for $K>2/9$ since the ground state changes from the 120$^\circ$ structure to the FQ state. Interestingly, one sees that the FQ state also sustains a topologically stable $Z_2$ vortex with a parity-like topological quantum number \cite{Mermin}. A typical example of such $Z_2$ vortex is illustrated in Fig.2b: It corresponds to a $\pi$ turn ($\pi$ disclination) of the director vector. 

Fig.1b exhibits the vorticity modulus for $K=0.5$. As can be seen from the figure, a vortex-induced topological transition occurs at $T_V\simeq 0.37$ in the temperature region where the FQ SRO has been developed. Here note the difference in the size dependences of $v$ and of $Q_F$ (or $m_f$) at low temperatures: With increasing $L$ at $T\lsim T_V$, while $v$ tends to increase slightly tending to a nonzero value, $Q_F$ or $m_f$ tends to decrease. Each size dependence corresponds to the LRO and the SRO, respectively.
%This indicates that the finite-temperature topological transition associated with the vortex condensation persists for all positive values of $K$.  Indeed, our MC simulations performed for other values of $K>0$ support this conjecture, whereas for certain intermediate values of $K$ the model turns out to exhibit successive transitions. Further details will be reported elsewhere.

\medskip\noindent
{\it Antiferroquadrapolar case $K<0$\/}

In the case of the AFQ coupling $K<0$, the ground state of three spins on a triangle remains to be a 120$^\circ$ spin structure for $K> -1$, whereas for $K<-1$ it takes a non-coplanar structure with an angle between two spins $\theta$ equal to $\cos \theta=1/(2K)$.
%, which, in the $K\rightarrow -\infty$ limit tends to 90. 
The change in the spin configuration at $K=-1$ is continuous. For $K<-1$, due to the non-coplanarity of the spin structure, the ground state possesses two distinct ``chiral'' states with mutually opposite signs of the scalar chirality $\vec S_a\cdot \vec S_b\times \vec S_c$. 
%: If the three spin vectors $\vec S_a$,¡¡$\vec S_b$ and $\vec S_c$ form the ground state, there is always another set of spin vectors ($\vec S_a$,$\vec S_b$,$\vec S'_c$), ($\vec S_a$,$\vec S'_b$,$\vec S_c$) and ($\vec S'_a$,$\vec S_b$,$\vec S_c$) with an opposite local chirality, which also form the ground state. 
This local chiral degeneracy has important consequence on the property of an infinite lattice, as
%By making use of this local degeneracy, the three sublattice periodicity inherent to the 120$^\circ$ structure can be easily lost: {\it e.g.\/} even if one starts with the three spins ($\vec S_a$,$\vec S_b$,$\vec S_c$) at a given triangle, one may get another set  ($\vec S'_a$,$\vec S_b$,$\vec S_c$) at the neighboring triangle, then ($\vec S'_a$,$\vec S_b$,$\vec S''_c$) at further neighboring triangle, {\it etc\/}. 
the sign of the local chirality tends to take random spatial pattern in the ground state, destroying the three-sublattice AF LRO. As we shall see below, such a ground state still can sustain the AFQ order with the three-sublattice periodicity.

%Generically, for an infinite lattice, one gets infinitely many spin configurations except for some special values of $K$. An interesting observation here is that the spin configurations for each triangle generated in such a way are arranged in the form of three sheets (unit circle) in three-spin space as illustrated in Fig.*. In fact, each sublattice a$\sim$c forms one sheet, each of which makes an angle equal to 60 with the other. Hence, the ground state realized here is the so-called antiferroquadrapolar (AFQ) or antiferro-nematic state with three-sublattice periodicity. Note that the type of AFQ state realized in our classical model differs somewhat from the one analyzed in Ref.* for the quantum $S=1$ bilinea-biquadratic model. 

In Fig.4a, we show a typical snapshot of spin directions observed at a temperature $T=0.17$, where a typical three-sublattice AFQ pattern is realized,  with the A-, B- and C-sublattice spins pointing to, say, $\pm S_x$, $\pm S_y$ and $\pm S_z$ directions with equal probability. Since such a locally orthogonal spin structure is not a ground state for $|K|<\infty$, its stabilization should be an entropic effect. 
%At lower temperature where the energy dominates, the orthogonal AFQ state should be destablized. 
In Fig.4b, we show a typical snapshot of spin directions at a lower temperature $T=0.01$, where a non-orthogonal AFQ state is realized in which spins on each triangle locally satisfy the above-mentioned ground-sate condition. 

% In quantum magnets, quantum fluctuations might stabilize the orthogonal AFQ state to lower temperatures, even down to $T=0$.

%
\begin{figure}[ht]
\begin{center}
\includegraphics[scale=0.25,angle=0]{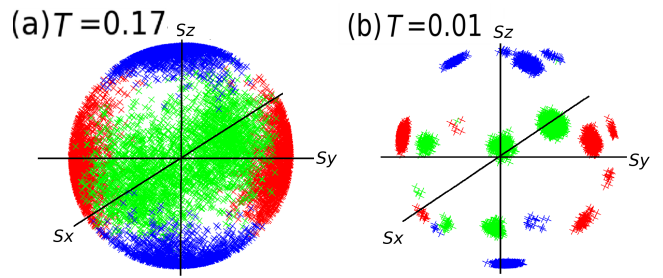}
\end{center}
\caption{
Snapshots of spin directions mapped onto a unit sphere in spin space for $K=-3$, at temperatures (a) $T=0.17$ and (b) $T=0.01$. Each color represents each sublattice.
}
\end{figure}

In Fig.3b, we show $m_f$ and the AFQ order parameter $Q_{AF}$ defined by
\begin{equation}
(Q_{AF})^2=3\sum_{\mu,\nu=x,y,z}<(\frac{1}{N}\sum_iq_{i\mu\nu}e^{i\vec q\cdot \vec r_i})^2>,
\end{equation}
 for the case of $K=-3$. The AFQ SRO turns out to develop rather sharply at $T\simeq 0.55$, where the standard AF SRO is still kept small. The AF SRO grows at a lower temperature $T\simeq 0.15$. The orthogonal AFQ spin structure illustrated in Fig.4a is realized in the intermediate temperature range $0.55\gsim T \gsim 0.15$, whereas the non-orthogonal AFQ state illustrated in Fig.4b is realized in the lower temperature range $T \lsim 0.15$. 
%The size dependence of $m_f$ suggests that $m_f$ vanishes in the thermodynamic limit even at $T\rightarrow 0$.
%The Z$_2$ vortex based on the noncollinear AF order is expected to survive at least up to $K=-1$, 

In the AFQ state, the order-parameter space is isomorphic to that of biaxial nematics. The topological defect structure of biaxial nematics has been analyzed \cite{Mermin}: It sustains a vortex whose topological quantum number is given by the quarternion group, or more precisely, its five conjugacy classes. In addition to the vortex-free state, there are four topologically distinct vortices. Interestingly, the quarternion group is non-Abelian, which might lead to a glassy dynamics via a peculiar combination rule of vortices. Even in such an exotic case,  the vortex binding-unbinding mechanism is expected to operate, {\it i.e.\/}, one expects a vortex-induced topological transition.

Fig.1c exhibits the vorticity modulus for $K=-3$. As can be seen from the figure, a vortex-induced topological transition takes place at $T_V\simeq 0.5$ in the temperature region where the AFQ SRO order is developed but the magnetic SRO is kept suppressed. In contrast to the $K>0$ case, the vorticity modulus exhibits a second anomaly around a temperature $T_2\simeq 0.15$ considerably lower than the vortex transition temperature. 
%In fact, around $T_2$, the standard magnetic SRO grows somewhat: See also the behavior of $m_F$ shown in Fig.*. 
Details of this second transition (or crossover) remains to be elucidated.

\medskip\noindent
{\it Implications to NiGa$_2$S$_4$\/}

 Based on our finding that the bilinear-biquadratic triangular Heisenberg AF exhibits a vortex-induced topological transition, we wish to discuss its possible implications to NiGa$_2$S$_4$. We argue that the experimentally observed ``transition'' of NiGa$_2$S$_4$ might be originated from a vortex-induced topological transition. The relevant vortices could be (i)  $Z_2$ vortices based on the noncollinear AF order for smaller $|K|$, (ii) $Z_2$ vortices based on the FQ order for largely positive $K$, and (iii) quarternion vortices based on the AFQ order for largely negative $K$. Whichever situation (i)-(iii) applies, the scenario immediately explains the experimental observation that the spin correlation length remains finite even at and below the transition. The specific heat is expected to show no appreciable anomaly at the transition, only a rounded peak above it, which seems consistent with experiments. Recent experiments on the nonmagnetic impurity effect have revealed that, as the impurity concentration is reduced toward the pure limit, the extent of the spin-glass-like hysteretic behavior is suppressed, while the freezing temperature $T_f$ itself {\it increases} \cite{Nambu}. This observation is also consistent with our view that the topological transition intrinsic to the pure system induces a spin-glass-like freezing in the corresponding impure system.

 The next question is obviously which type of vortex is relevant in NiGa$_2$S$_4$. Very recent NQR and $\mu$SR measurements indicate that static internal fields set in below $T_f$ accompanied with a divergent increase of the correlation time toward $T_f$, at least within experimental time window \cite{Ishida}. This observation of internal fields appears compatible only with the case (i) above. 
% While further studies are desirable to determine which type of vortex plays a role in NiGa$_2$S$_4$, we feel the case (i) is most likely with some possibility left to the case (iii). 
In the case (i), the low-temperature phase should be dominated by spin-wave excitations: It is a near critical phase characterized by large but still {\it finite\/} spin correlation length and correlation time. Then, spin waves would be responsible for the $T^2$ specific heat. Indeed, 
%
%Experimentally,  the $T^2$ behavior of the specific heat tends to set in around the transition temperature $T_V$, which is consistent with such a picture. 
%
Fujimoto recently accounted for the $T^2$ specific heat based on the spin-wave excitations of the noncollinear AF order of the $S=1$ quantum magnets, neglecting the vortex degrees of freedom \cite{Fujimoto}. Vortex-free assumption of Ref.\cite{Fujimoto} is well justified at $T<T_V$, if there occurs a topological transition.  Note that, in this vortex scenario, the correlation time does not truly diverge at $T_V$(=$T_f$), but only grows sharply at $T_V$ exceeding the experimental time scale, and stays long in a wide temperature range below $T_V$. Such a near critical behavior realized below $T_V$ seems consistent with the NQR observation \cite{Ishida}. One may suspect that a weak interplane coupling $J'$, which should exist in real NiGa$_2$S$_4$, inevitably induces the 3D AF LRO immediately below $T_V$. However, this is not necessarily the case: If $J'$ is sufficiently small satisfying $J'\xi(T_V) ^2 \lsim k_BT_V$, $\xi(T_V)$ being the spin correlation length at $T_V$, the 3D AF LRO needs not set in even below $T_V$. Finiteness of $\xi$ and smallness of $J'$ are essential in preventing the vortex ordered state from forming the 3D AF LRO. At still lower temperatures, $\xi$ diverges exponentially toward $T=0$, eventually leading to the onset of the magnetic LRO at a certain temperature $T'<T_V$. 

In NiGa$_2$S$_4$, distant neighbor interactions neglected in the present analysis, particularly the third-neighbor interaction,  compete with the nearest-neighbor one leading to an incommensurate spin structure at low temperature \cite{Nakatsuji}. We note that the vortex transition discussed here is not specific to the 120$^\circ$ spin structure realized in the nearest-neighbor model, but generically expected for the noncollinear spin order including the incommensurate one, although details of the transition needs to be clarified further. 
%Of course, if $J'$ is stronger than the above threshold, the 3D AF LRO will be formed at $T_V$. 
The vortex scenario might also apply to the S=3/2 triangular AF NiCrO$_2$ \cite{Olariu}.

Finally, the noncollinear AF order 
%underlying the vortex 
might explain another noticeable feature of experiments that the $T^2$ specific heat is quite robust against applied magnetic fields \cite{Nakatsuji}.  It is because the noncollinear AF ground state in magnetic fields is capable of keeping an accidental degeneracy not related to the Hamiltonian symmetry, essentially of the same amount as in the zero-field case \cite{KawaMiya2}. Hence, at the classical level, this accidental degeneracy gives rise to {\it pseudo\/}-Goldstone modes even in applied fields, which may account for the robustness of the low-temperature specific heat, while this degeneracy would become approximate in quantum systems.

%The only potential probelm might be that the magnetic correlation length deduced from neutron measurements appears to stay shortl even at lower temperature $T<<T_V$, while in this scenario it should eventually diverge toward $T=0$.

%Concerning the nematic scenario,  the real microscopic mechanism to induce a large $K$-value required to stabilize the spin nematic order remains to be clarified. In the FQ case ($K>0$),  Ref.* argued that the quasi-degeneracy of orbitals might cause an effective enhancement of the $K$-value \cite{Lauchli}. A possible drawback of the FQ scenario might be the observed robustness of the low-temperature specific heat against applied magnetic fields, which might be difficult to explain in the FQ scenario since the FQ order inevitably selects a unique axis in spin space. In the AFQ case ($K<0$), it remains to be seen whether the accidental-degeneracy mechanism works.

%Both in the FQ and AFQ cases, at least at the classical level, the application of even weak fields tends to supress the local degeneracy of the ground state and induce the standard magnetic long-range order at $T=0$. Of course, the quantum effect totally neglected here might change these conclusions which remain to be seen.

In summary, we studied the ordering properties of the AF Heisenberg model on the triangular lattice with the the bilinear-biquadratic coupling, and have shown that the model exhibits a vortex-induced topological transition. The relevant vortices could be of three different types, depending on the value of the biquadratic coupling. It was then suggested that the peculiar phase transition recently observed in NiGa$_2$S$_4$ might have its origin in such a vortex-induced topological transition. 

%Discussing the three possible patterns of the vortex order, we have argued that the most likely candidate for NiGa$_2$S$_4$ might be the $Z_2$ vortex based on the noncollinear magnetic orderd state. On the basis of this picture, we also presented an accidental-degeneracy mechanism to account for the observed robustness of the low-temperature specific heat against applied magnetic fields.

The authors thank S.Nakatsuji, K.Ishida, Y.Nambu, H.Tsunetsugu, M.Arikawa, S.Fujimoto for discussion.

\end{document}